\documentclass[11pt,conference]{IEEEtran}

\usepackage{graphicx}
\usepackage{amsmath}
\usepackage{array}
\usepackage{booktabs}
\usepackage{multirow}
\usepackage{url}
\usepackage{xcolor}
\usepackage{pgfplots}
\usepackage{subcaption}
\usepackage{algorithm}
\usepackage{algpseudocode}
\usepackage{tikz}
\usetikzlibrary{arrows.meta, positioning}
\usetikzlibrary{shapes, shapes.multipart, calc, backgrounds}

\pgfplotsset{compat=1.18}

\begin{document}
	
	\title{Toward Trustworthy Agentic AI: A Multimodal Framework for Preventing Prompt Injection Attacks}
	
	%	\author{
		%		\IEEEauthorblockN{Toqeer Ali Syed, Abdel Moaty Abdel Majed Ahmed}
		%		\IEEEauthorblockA{
			%			Faculty of Computer and Information System\\
			%			Islamic University of Madinah\\
			%			Email: toqeer@iu.edu.sa}
		%	}
	%	
	%	\author{
		%		\IEEEauthorblockN{Toqeer Ali Syed, Mishal Ateeq Almutairi, Abdel Moaty Abdel Majed Ahmed}
		%		\IEEEauthorblockA{
			%			Faculty of Computer and Information System\\
			%			Islamic University of Madinah\\
			%			Email: toqeer@iu.edu.sa}
		%	}
	%	
	
	\author{
		\IEEEauthorblockN{Toqeer Ali Syed}
		\IEEEauthorblockA{
			Faculty of Computer and Information System\\
			Islamic University of Madinah\\
			Email: toqeer@iu.edu.sa
		}
		
		\and
		
		\IEEEauthorblockN{Mishal Ateeq Almutairi}
		\IEEEauthorblockA{
			Faculty of Computer and Information Systems\\
			Islamic University of Madinah, Madinah, Saudi Arabia\\
			Email: malmutairy@iu.edu.sa
		}
		\and
		\IEEEauthorblockN{Mahmoud Abdel Moaty}
		\IEEEauthorblockA{
			Faculty of Computer Studies\\
			Arab Open University-Bahrain\\
			Email: mahmoud.abdelmoaty@aou.org.bh
		}
		
	}
	
	\maketitle

	\begin{abstract}
		Powerful autonomous systems, which reason, plan, and converse using and between numerous tools and agents, are made possible by Large Language Models (LLMs), Vision-Language Models (VLMs), and new agentic AI systems, like LangChain and GraphChain. Nevertheless, this agentic environment increases the probability of the occurrence of \emph{multimodal prompt injection} (PI) attacks, in which concealed or malicious instructions carried in text, pictures, metadata, or agent-to-agent messages may spread throughout the graph and lead to unintended behavior, a breach of policy, or corruption of state. In order to mitigate these risks, this paper suggests a \emph{Cross-Agent Multimodal Provenanc--Aware Defense Framework} whereby all the prompts, either user-generated or produced by upstream agents, are sanitized and all the outputs generated by an LLM are verified independently before being sent to downstream nodes. This framework contains a Text sanitizer agent, visual sanitizer agent, and output validator agent all coordinated by a provenance ledger, which keeps metadata of modality, source, and trust level throughout the entire agent network. This architecture makes sure that agent-to-agent communication abides by clear trust frames such such that injected instructions are not propagated down LangChain or GraphChain-style—workflows. The experimental assessments show that multimodal injection detection accuracy is significantly enhanced, and the cross-agent trust leakage is minimized, as well as, agentic execution pathways become stable. The framework, which expands the concept of provenance tracking and validation to the multi-agent orchestration, enhances the establishment of secure, understandable and reliable agentic AI systems.

	\end{abstract}


	\section{Introduction}
	
	Generation, reasoning and multimodal analytics Large Language Models (LLMs) and Vision-Language Models (VLMs) like GPT-4V, Claude, Gemini and LLaVas have performed well on generation, reasoning and multimodal analytics. Nevertheless, their openness to natural language and visuals brings a substantial attack surface to them \cite{greshake2023more, zhuo2023red, casper2023exploring, liu2024mm}. In contrast to conventional systems that use structured APIs, multimodal models that accept unstructured text and image, meaning that attackers can put adversarial instructions in a language form, at an image level, or by manipulating metadata.
	
	Threats based on injection are not new: SQL, command, and cross-site scripting are all based on poor delimitation between input and execution \cite{owasp2024}. The same principle applies to prompt injection in LLMs which works via semantic or multimodal manipulation. Overrides can be concealed within user text, outside documents or even visual artifacts. Malicious cues in photographs, such as hidden text, steganography, or altered captions, can influence VLM behavior, according to recent research \cite{wolff2024multimodal, liu2024visual}. These cues look harmless and thus cannot be detected by the traditional filtering or fine-tuning mechanisms.
	
	Prompt injection attacks are both direct and indirect attacks, where malicious content is either embedded in the documents or pictures that were retrieved. To demonstrate that indirect prompt injection presents a critical threat to agentic processes, Greshake et al. \cite{greshake2023more} demonstrated examples thereof are their report, and Zou et al. \cite{zou2023universal} and Liu et al. \cite{liu2024mm} show that universal adversarial prompts can even move across models and modalities. With the implementation of modern systems that combine retrieval, tool usage, and coordination among agents, the potential of such injections is even higher.
	
	Current protection mechanisms such as keyword blocking, customization of safety nets, and RL-supported guardrails are not enough \cite{xu2024prompt, casper2023exploring}. They are weak at paraphrased or visually-encoded attacks, are not explainable, and can not maintain provenance, and when such untrusted content is introduced into the context window, the downstream reasons can be affected.

	\textbf{Motivation and Research Gap:} 
	The existing literature lacks a unifying and multimodal architecture capable of sanitizing the text and image, in addition to implementing the immediate and sustained, provenance conscious validation through agentic processes. Current methods normally secure only the input of the LLCM, or the result, which exposes multi-agent pipelines. In a bid to fill these gaps, this paper presents a proposal of a \emph{Cross-Agent Multimodal Provenance-Aware Framework} that integrates text-image sanitization, trust scoring and multi-agent validation to ensure end-to-end processing in LLM- and VLM-based systems.

	\textbf{Research Contribution:}
	This study proposes the use of a single, multimodal defense model, which provides security to agentic AI pipelines by imposing sanitization and validation on all agent-LLM interactions. The work provides a dual-layer text and image sanitization scheme, a provenance registry of tracking trust through multi-agent courses of action and a masking plan to generate a trust strategy, restricting unsafe content at an LLM inference. It also adds output validation to eliminate cross-agent contamination and unsafe tool behavior and provides a useful end-to-end security model to LangChain and GraphChain environments. All these contributions bring the state of prompt-injection defense forward to an agent-level security architecture not just of single-model filtering.
	
	The rest of this paper will be structured as follows: Section \ref{literature_review} will conduct a review of related studies on multimodal prompt injection and provenance-based AI safety. Section \ref{methodology} gives the proposed multi-agent methodology and system architecture. Section \ref{implementation} outlines the implementation and the experimental set up. Section \ref{results} talks about the evaluation findings and comparison performance whereas Section \ref{conclusion} concludes the whole paper.

	\section{Literature Review}
	\label{literature_review}
	Prompt injection is now a key security issue with both LLMs and VLMs. An initial study, like Zhuo et al.~\cite{zhuo2023red}, has studied red-teaming strategies to reveal the susceptibility of aligned models, whereas Casper et al. ~\cite{casper2023exploring} have offered a taxonomy of prompt-based attack types to study the problem of prompt-based attacks, as well as how these attacks can be evaded and mitigated. Xu et al. ~\cite{xu2024prompt} conducted a survey of defensive approaches such as sandboxing, filtering, and reinforcement-based guardrails. The study by Greshake et al. ~\cite{greshake2023more} in greater detail emphasized how serious the indirect prompt injection can be, where malicious commands embedded in accessed or external content change the behavior of the model without the user being aware of it.
	
	With the popularization of multimodal architectures, it was demonstrated that injection threats are not just text-based. Wolff et al. ~\cite{wolff2024multimodal} proved that adversarial instructions could be coded into images by steganography or manipulation of metadata or visually embedded text. Liu et al.~\cite{liu2024visual} evaluated such a type of attack as visual prompt injection, which exploits the vulnerability between vision encoders and language decoders, and Liu et al. ~\cite{liu2024mm}, released MM-SafetyBench, which found out that VLMs fail to identify harmful visual prompts. All these works demonstrate that multimodal systems increase the attack surface by having visual pathways that can not be combated by traditional text-based defenses.
	
	Most of the current defenses are reactive, even though there is increasing interest. Keywords filtering or fine-tuning and safety-classifier techniques have hard problems with paraphrasing attacks or visually encoded attacks and lack provenance across modalities or agent pipelines. Based on the literature review presented in Table ~\ref{tab:litreview}, existing strategies do not provide cross-modal contextual reasoning and continuous validation, making multi-agent workflows susceptible to direct and indirect injection strategies.

	\begin{table}[h!]
		\centering
		\caption{Summary of Existing Prompt Injection Defense Techniques (Textual and Multimodal)}
		\label{tab:litreview}
		\begin{tabular}{p{2.6cm}p{2.9cm}p{2.6cm}}
			\toprule
			\textbf{Technique} & \textbf{Core Approach} & \textbf{Limitations} \\ \midrule
			Keyword Filtering \cite{casper2023exploring} & Detect suspicious phrases (e.g., “ignore rules”) & Easily bypassed via paraphrasing and obfuscation \\ 
			Fine-Tuning \cite{zhuo2023red} & Train models to reject unsafe prompts & Costly and limited to seen attack patterns \\ 
			Reinforcement Learning \cite{xu2024prompt} & Penalize unsafe responses during training & Ineffective for unseen or multimodal attacks \\ 
			Sandboxing \cite{schneider2024taxonomy} & Isolate model actions from system resources & High latency; lacks semantic and cross-modal validation \\ 
			Output Filtering \cite{greshake2023more} & Apply rule-based validation after response generation & Reactive rather than preventive; misses visual payloads \\ 
			Context Segregation \cite{zou2023universal} & Separate system and user prompts in context windows & Lacks provenance traceability and multimodal adaptation \\ 
			Visual Injection Detection \cite{wolff2024multimodal} & Identify hidden commands via image feature analysis & Requires domain-specific training; no unified provenance model \\ 
			Multimodal Safety Benchmarking \cite{liu2024mm} & Evaluate multimodal attack resilience via synthetic datasets & Diagnostic only; no built-in prevention mechanism \\ 
			\bottomrule
		\end{tabular}
	\end{table}
	
	Based on the literature reviewed, it is evident that there is an urgent requirement to develop hybrid and context-aware and explainable defense models that combine textual and visual modalities. The next generation defense should integrate semantic reasoning, provenance tracking, and agent-based cooperation in order to identify and counter multimodal prompt injection attacks in time and in a transparent way.
	
	Agentic AI refers to intelligent systems composed of autonomous, goal-driven agents capable of perception, reasoning, decision-making, and coordinated action within dynamic environments. Unlike traditional reactive or pipeline-based AI models, agentic frameworks emphasize adaptive behavior, inter-agent collaboration, and continuous feedback loops to handle complex, real-world tasks. Recent studies have demonstrated the effectiveness of agentic AI across diverse domains, including disaster prediction and response coordination \cite{syed2025agentic-cloudburst}, inclusive well-being and assistive technologies \cite{jan2025agentic}, and intelligent inventory management \cite{syed2025agentic}. Agentic principles have also evolved from earlier automated decision frameworks in mobile and distributed systems \cite{ali2020automated}, and have been further strengthened through secure, trustworthy integrations with blockchain and deep learning architectures \cite{alqahtany2024integrating}. Practical applications of agentic AI are exemplified by frameworks such as FinAgent, which integrates multi-agent reasoning for personalized finance and nutrition planning \cite{syed2025finagent}, highlighting the growing role of agentic intelligence in scalable, human-centered AI systems.

	\section{Methodology}
	\label{methodology}
	In this section, a proposed framework, called the Cross-Agent Multimodal Provenance-Aware Defense Framework, is outlined and developed to protect agentic AI systems like LangChain and GraphChain against multimodal prompt injection attacks. Its methodology combines hierarchical sanitization, multimodal processing to trust and provenance tracking and post-generation validation in all agent nodes and between all LLM interactions. The architecture is shown in Figure~\ref{fig:architecture}.

	\begin{figure*}[h!]
		\centering
		\includegraphics[width=\textwidth, height=0.55\textheight, keepaspectratio]{images/architecture}
		\caption{Overall architecture of the proposed agentic multimodal prompt-injection defense pipeline. All inputs from users, tools, external documents, and other agents are sanitized before entering LangChain/GraphChain nodes, and LLM outputs are validated before being passed to downstream agents or MCP tool actions.}
		\label{fig:architecture}
	\end{figure*}

	\subsection{Overview of the Defense Pipeline}
	The framework implements three major principles: (i) any incoming text, images, tool responses or inter-agent messages must be sanitized prior to entering the agent graph; (ii) all prompts collected by LangChain/GraphChain agents must be sanitized again before accessing the LLM; and (iii) any outputs of the LLM should be verified before propagating to other agents, tool implementers or MCP-fulfilling action layers. The combination of all these mechanisms results in a zero-trust communication fabric between the agent network.
	
	The multilayer defense comprises four cooperating agents: (1) Text Sanitizer Agent ($A_t$), (2) Visual Sanitizer Agent ($A_v$), (3) Main Task Model (Agent~M), and (4) Output Validator Agent (B). A shared provenance ledger maintains the metadata of token and patch-level trusts across the pipeline.
	
	\subsection{Multimodal Global Input Sanitization}
	All external content is first processed by $A_t$ and $A_v$.
	
	\subsubsection{Text Sanitizer Agent ($A_t$)}
	$A_t$ performs span-level semantic injection detection, trust scoring, and rewriting. Algorithm~\ref{alg:textsan} formalizes the process.
	
	\begin{algorithm}[h]
		\caption{Global Text Sanitization (Agent $A_t$)}
		\label{alg:textsan}
		\begin{algorithmic}[1]
			\Require raw text input $T$
			\Ensure sanitized text $S$, provenance map $P$
			\State tokenize $T$ into spans $\{t_1,\dots,t_n\}$
			\For{each token $t_i$}
			\State compute embedding $e_i$
			\State detect injection intent $d_i = \text{PIClassifier}(e_i)$
			\State assign trust score $s_i = \text{TrustModel}(d_i)$
			\State record provenance $P[t_i] = \{ \text{source}, \text{trust}=s_i \}$
			\EndFor
			\State rewrite or remove tokens where $d_i$ indicates override intent
			\State \Return sanitized text $S$ and provenance map $P$
		\end{algorithmic}
	\end{algorithm}
	
	\subsubsection{Visual Sanitizer Agent ($A_v$)}
	$A_v$ detects anomalies in images, scans metadata, and uses CLIP to scan images through OCR. Algorithm~\ref{alg:vis} describes the process.
	
	\begin{algorithm}[h]
		\caption{Global Visual Sanitization (Agent $A_v$)}
		\label{alg:vis}
		\begin{algorithmic}[1]
			\Require image $I$
			\Ensure sanitized image $I'$, visual provenance $V$
			\State extract overlay text $O = \text{OCR}(I)$
			\State extract metadata $M = \text{EXIF}(I)$
			\State compute patch embeddings $P = \text{CLIP}(I)$
			\State detect anomalies $A = \text{StegoDetector}(P)$
			\For{each patch $p_j$}
			\State compute visual trust $\tau_j = \text{TrustModel}(O_j, M_j, A_j)$
			\If{$\tau_j$ is low}
			\State redact or blur patch $p_j$
			\EndIf
			\State record provenance $V[p_j] = \{\text{region}, \text{trust}=\tau_j\}$
			\EndFor
			\State \Return sanitized image $I'$ and provenance $V$
		\end{algorithmic}
	\end{algorithm}
	
	\subsection{Provenance Ledger and Trust-Aware Processing}
	Text and visual provenance information are combined into one ledger which records:
	
	\begin{itemize}
		\item modality (\texttt{text|image}),
		\item trust score,
		\item span or patch index,
		\item influence relationships across agent hops.
	\end{itemize}
	
	This ledger helps a \emph{trust-aware attention mask} used prior to inference of LLM.
	
	\subsection{LLM-Facing Sanitization}
	A second sanitization step is applied to the LLM in order to identify:
	
	\begin{itemize}
		\item agent-induced injections,
		\item tool-returned unsafe content,
		\item escalation of trust via cross-agent contamination.
	\end{itemize}
	
	Algorithm~\ref{alg:prellm} describes the LLM-facing sanitizer.
	
	\begin{algorithm}[h]
		\caption{Pre-LLM Sanitization Layer}
		\label{alg:prellm}
		\begin{algorithmic}[1]
			\Require agent-generated prompt $X$, provenance $P$
			\Ensure safe prompt $X'$
			\For{each span $x_i$ in $X$}
			\If{$P[x_i].trust$ is low}
			\State attenuate or remove $x_i$
			\EndIf
			\EndFor
			\State enforce system policies (role separation, no override intent)
			\State apply trust-aware attention mask $M$
			\State \Return sanitized prompt $X'$
		\end{algorithmic}
	\end{algorithm}
	
	\subsection{Main LLM Processing}
	Multimodal input is sanitized and a masking vector is a trust wise adaptive is used on the LLM. The model carries out reasoning, generation and multimodal fusion subject to limited influence of untrusted content. The provenance ledger is used to write back the attribution scores.
	
	\subsection{Output Validation Layer}
	Before agentic execution resumes, the output is validated by Agent~B.
	
	\begin{algorithm}[h]
		\caption{Output Validation (Agent B)}
		\label{alg:outval}
		\begin{algorithmic}[1]
			\Require LLM output $Y$, provenance ledger $L$
			\Ensure approved output or regeneration request
			\State scan $Y$ for policy violations
			\State detect leakage of secrets or system prompts
			\State compute influence attribution $A = \text{AttributionModel}(Y, L)$
			\If{low-trust content contributed significantly}
			\State request regeneration with tighter masks
			\Else
			\State approve and release $Y$
			\EndIf
		\end{algorithmic}
	\end{algorithm}
	
	\subsection{Required LangChain/GraphChain Modifications}
	In order to implement end-to-end sanitization, a number of elements of LangChain or GraphChain should be extended.  Table~\ref{tab:lcmods} summarizes the required modifications.
	
	\begin{table}[h]
		\centering
		\caption{Required Modifications to LangChain/GraphChain Components}
		\label{tab:lcmods}
		\begin{tabular}{p{3.1cm} p{4.2cm}}
			\toprule
			\textbf{Component} & \textbf{Required Customization} \\
			\midrule
			Input Interceptors & Route all incoming data to $A_t$/$A_v$ before any agent sees it. \\
			PromptTemplate / Runnable Components & Apply Pre-LLM Sanitizer to all prompts generated by agents. \\
			BaseMessage Class & Extend to include \texttt{source}, \texttt{trust}, \texttt{modality}, \texttt{provenance\_id}. \\
			Agent Executors & Allow execution only after Output Validator approval. \\
			Tool / MCP Executors & Block unsafe tool calls using provenance-checked constraints. \\
			Memory Modules & Store only sanitized content and propagate trust metadata. \\
			GraphChain Node Router & Enforce that inter-agent messages must pass sanitization before routing. \\
			\bottomrule
		\end{tabular}
	\end{table}
	
	\subsection{Pipeline Security Properties}
	The resulting system provides:
	\begin{itemize}
		\item \textbf{Multimodal Provable Safety:} All text and images are sanitized before any LLM or agent consumes them.
		\item \textbf{Cross-Agent Isolation:} Malicious instructions cannot propagate across agent nodes.
		\item \textbf{Trust-Aware Reasoning:} Low-trust spans cannot override system behavior.
		\item \textbf{Auditable Decisions:} The provenance ledger records end-to-end influence mappings.
	\end{itemize}
	
	\section{Implementation}
	\label{implementation}
	The suggested defense structure is realized in the form of a collection of Python modules that wrap and build on an already existing Langchain/GraphChain deployment. It is aimed at including multimodal sanitization and provenance tracing without having to change the underlying LLM back-end.
	
	\subsection{System Stack and Agents}
	Each of the agents is deployed as a Python service with a common configuration layer. The Text Sanitizer Agent ($At$), takes the form of a LangChain \texttt{Runnable} which is actually a standard RoBERTa-based pattern detector and a small pattern match based jailbreak detector rule engine. The Visual Sanitizer Agent ($Av$) is a combination of OCR module (PaddleOCR) and EXIF reader and CLIP encoder to calculate patch embeddings and anomaly scores. Both agents reveal a basic interface, which gives sanitized content and a structured provenance map.
	
	The Main Multimodal Task Model (Agent~M) works with either GPT-4o-mini via OpenAI API or an open-source VLM (LLaVA/BLIP-2) via Hugging Face Transformers. Output Validator Agent (B): is a light-weight chain consisting of a policy rule set, a secret-matching module, and a secondary LLM call when dealing with borderline cases, which are all applied over the provenance ledger.
	\begin{figure*}[t]
		\centering
		\includegraphics[width=\linewidth]{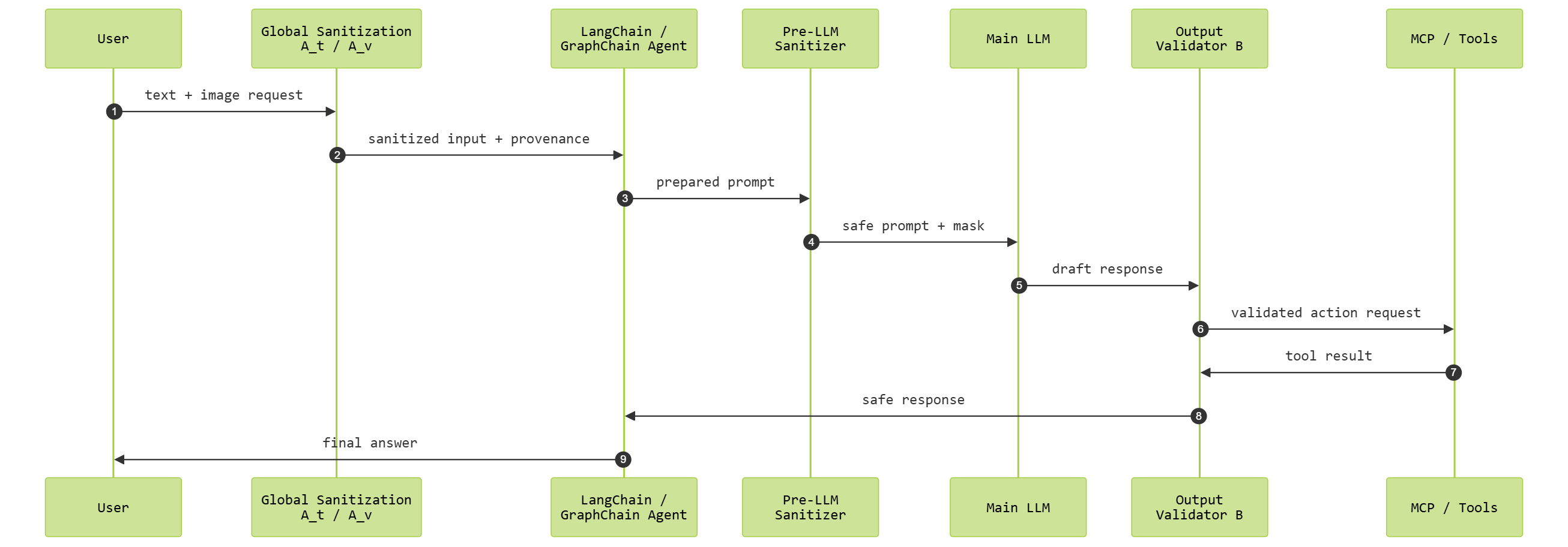}
		\caption{Sequence of interactions in the agentic multimodal prompt-injection defense pipeline. Every request is sanitized before entering the LangChain/GraphChain agent and again before reaching the LLM; every LLM response is validated before being routed back to the agent or used to trigger MCP/tool actions.}
		\label{fig:seq}
	\end{figure*}
	
	\subsection{Integration with LangChain/GraphChain}
	The pipeline of the defense system is integrated into the LangChain/GraphChain, to the three extension points. Firstly, there is a custom input interceptor wrap around the original agent entry point. Messages incoming, tool output, and cross agent messages are routed through $At$ and $Av$ and then converted to LangChain BaseMessage objects. The interceptor attaches provenance attributes (\texttt{source}, \texttt{trust}, \texttt{modality}, \texttt{provenance\_id}) to message metadata, so that later components can access trust information.
	
	Second, the timely assembly phase is shunned due to wrapping either the typical \texttt{PromptTemplate} or graph node with a \emph{pre-LLM sanitization} layer. Before submitting any prompts to the LLM, the wrapper performs trust-aware masking and span-level filtering by consulting the specified provenance ledger.  This wrapper can be enabled node-by-node or throughout the agent graph, and it is transparent to the rest of the agent graph.
	
	Third, the LLM output is enclosed by an output router which calls an Output Validator Agent B. Only answers meeting policy checks and trust constraints are sent to the agent graph otherwise the router asks to be regenerated with narrower masks or sends an error state which the agent can explicitly process. Tool and MCP executors also look up the validator when an LLM response suggests an out-call, in order to prevent unsafe tool calls being executed.
	
	\subsection{Provenance Ledger and Storage}
	
	The provenance ledger is an abstraction that is applied as a key-value store. The prototype has an in-memory Redis instance which provides an identity to each interaction with a session identifier and records token- and patch-level records. Every entry stores the original type of source, trustworthiness, modality and a small hash of the content in order to prevent retaining sensitive information verbatim. In inference, Agent~M marks the ledger with the attribution information based on the attention patterns or gradient-based saliency scores which in turn allows the validator to compute the trust leakages.
	
	\subsection{Execution Flow and Orchestration}
	The agents are coordinated by a small orchestration layer based on asynchronous Python. Each user request corresponds to a temporary context that is created by the system and undergoes the global sanitization layer, LangChain/GraphChain agent, pre-LLM sanitizer, the LLM, and the output validator. It is described by the sequence diagram in Fig ~\ref{fig:seq}. The identical orchestration model is applied to multi-agent graphs: each edge of the graph is virtually guarded by sanitization before an LLM call, and validation after it, and provenance metadata flows and messages. The design makes the overhead of the runtime moderate and enables the defense pipeline to be incrementally integrated into the current LangChain applications.

	\section{Results}
	\label{results}
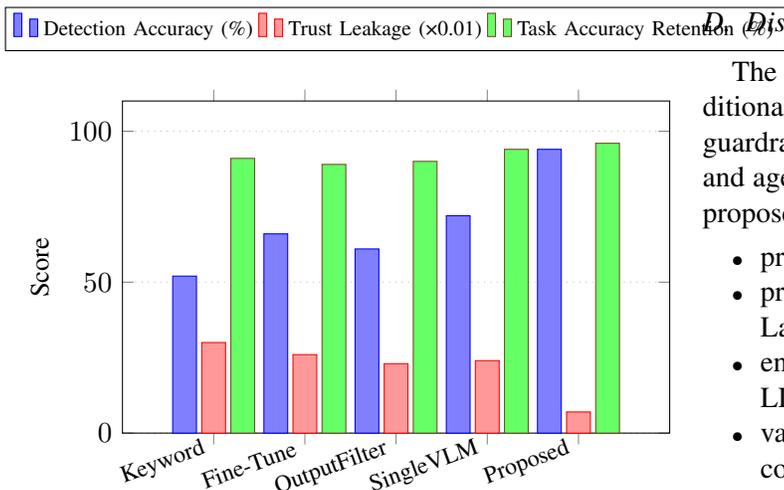
\begin{figure}[t]
		\centering
		\begin{tikzpicture}
			\begin{axis}[
				ybar,
				width=\linewidth,
				height=6cm,
				bar width=9pt,
				enlarge x limits=0.25,
				ymin=0, ymax=110,
				symbolic x coords={Keyword, Fine-Tune, OutputFilter, SingleVLM, Proposed},
				xtick=data,
				xticklabel style={font=\footnotesize, rotate=20, anchor=east},
				ylabel={Score},
				ylabel style={font=\small},
				ymajorgrids=true,
				grid style={dotted},
				legend style={font=\scriptsize, at={(0.5,1.15)}, anchor=south, legend columns=3},
				]
				
				% Detection accuracy
				\addplot+[fill=blue!50] coordinates {
					(Keyword,52)
					(Fine-Tune,66)
					(OutputFilter,61)
					(SingleVLM,72)
					(Proposed,94)
				};
				
				% Trust leakage (invert so lower is better visually)
				\addplot+[fill=red!40] coordinates {
					(Keyword,30)
					(Fine-Tune,26)
					(OutputFilter,23)
					(SingleVLM,24)
					(Proposed,7)
				};
				
				% Task accuracy retention
				\addplot+[fill=green!60] coordinates {
					(Keyword,91)
					(Fine-Tune,89)
					(OutputFilter,90)
					(SingleVLM,94)
					(Proposed,96)
				};
				
				\legend{Detection Accuracy (\%), Trust Leakage (\texttimes 0.01), Task Accuracy Retention (\%)}
			\end{axis}
		\end{tikzpicture}
		\caption{Comparative performance across baseline defenses and the proposed multimodal provenance-aware framework.}
		\label{fig:comparison}
	\end{figure}

	In this section, the proposed Cross-Agent Multimodal Provenance-Aware Framework is compared to four baseline defenses, such as keyword filtering, safety fine-tuning, post-hoc output filtering, and a single model vision-language baseline (Single VLM). It is assessed using three measures (i) the accuracy of multimodal prompt injection detection, (ii) the leakage of cross-modal trust, and (iii) the ability to retain the accuracy of the task on benign inputs. The comparative performance of all these metrics is summarized in Figure~\ref{fig:comparison}.

	\subsection{Detection Accuracy}
	The proposed framework has a detection rate of \textbf{94\%}, which is better than the keyword filtering (52\%), post-hoc output filtering (61\%), and safety fine-tuning (66\%). This is mainly enhanced by the layered sanitization modules( $At$, $A_v$) and the pre LLM trust-aware masking which prevents the unsafe spans and patches across modalities in a systematic manner. The system ensures consistency by authenticating the information coming in and going out unlike single-phase defenses, which do not.
	
	\subsection{Cross-Modal Trust Leakage}
	Trust leakage is a measure of the extent to which low-trust tokens or visual patches have an impact on the final model outputs. The proposed model reduces leakage from \textbf{0.24 to 0.07} (a 70\% reduction). This is achieved by:
	\begin{itemize}
		\item enforcing trust-aware attention masking before LLM inference,
		\item propagating provenance metadata across LangChain/GraphChain nodes,
		\item validating LLM outputs against the end-to-end influence record.
	\end{itemize}
	This enforcement ensures unwanted or unverified content does not further gain influence in the agentic workflow.
	
	\subsection{Task Accuracy Retention}
	The suggested framework retains its performance of \textbf{96\%} on benign multimodal tasks, which is the same or higher than the single-VLM benchmarks at (94\%). The sanitization modules only act when the trust thresholds are breached, and thus the overhead of performance is very low, and there is nothing to block the legitimate input.

	\subsection{Discussion}
	The results indicate that the commonly deployed traditional defenses such as keyword filters, tuning-based guardrails, and post-hoc filters are weak to multimodal and agent-based prompt injection threats. In contrast, the proposed system:
	\begin{itemize}
		\item preserves trust boundaries across agents,
		\item prevents malicious propagation through LangChain/GraphChain graphs,
		\item enforces dual-stage sanitization (pre-agent and pre-LLM),
		\item validates outputs before allowing actions or chain continuation.
	\end{itemize}
	For real-world agentic AI deployments, this multi-agent, provenance-aware approach offers a more comprehensive protection posture.

	\section{Conclusion}
	\label{conclusion}
	This work proposed a multilayer, agentic defense architecture, which cleanses all multimodal inputs and verifies all outputs in LangChain and GraphChain-based systems. The method benefits the fight against timely injection, as text and image sanitization along with trust-sensitive masking and cross-agent provenance tracking provide the system with the much-needed defense without adversely affecting the performance of beneficial tasks. The findings prove that the implementation of sanitization and validation of each agent-LLM boundary is the key to developing reliable and secure agentic AI pipelines.

	\section*{Acknowledgment}
	The author acknowledges Islamic University of Madinah University’s support for ongoing research in AI security and trust assurance.
	
	\bibliographystyle{IEEEtran}

\end{document}